# Temperature effects in the thermal conductivity of aligned amorphous polyethylene—A molecular dynamics study

Rajmohan Muthaiah, and Jivtesh Garg

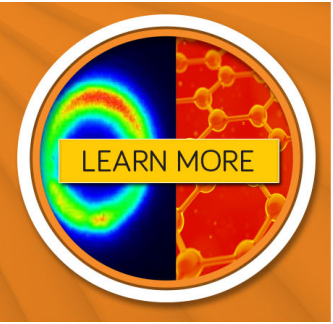

# Temperature effects in the thermal conductivity of aligned amorphous polyethylene—A molecular dynamics study

Rajmohan Muthaiah and Jivtesh Garg
*School of Aerospace and Mechanical Engineering, University of Oklahoma, Norman, Oklahoma 73019, USA*

We analyze, through molecular dynamics simulations, the temperature dependence of the thermal conductivity ($k$) of chain-oriented amorphous polyethylene (PE). We find that at increasing levels of orientation, the temperature corresponding to a peak $k$ progressively decreases. Un-oriented PE exhibits the peak $k$ at 350 K, while aligned PE under an applied strain of 400% shows a maximum at 100 K. This transition of peak $k$ to lower temperatures with increasing alignment is explained in terms of a crossover from disorder to anharmonicity dominated phonon transport in aligned polymers. Evidence for this crossover is achieved by manipulating the disorder in the polymer structure and studying the resulting change in temperature corresponding to peak $k$. Disorder is modified through a change in the dihedral parameters of the potential function, allowing a change in the relative fraction of *trans* and *gauche* transformations. The results shed light on the underlying thermal transport processes in aligned polymers and hold importance for low temperature applications of polymer materials in thermal management technologies. *Published by AIP Publishing.*
https://doi.org/10.1063/1.5041000

## I. INTRODUCTION

High thermal conductivity polymers play a key role in thermal management in a wide array of applications including water desalination,[1] solar energy harvesting,[2] automotive control units,[3] and micro-electronics.[4] Polymers offer several advantages relative to metals such as low weight, low cost, and lower fabrication energy.[5] Alignment of polymer chains has emerged as a promising approach to enhance the thermal conductivity of polymers.[6,7] Recently, $k$ of a single polyethylene (PE) nanofiber with highly aligned PE chains was measured to be 104 W/mK, almost 200 times[6] larger than the $k$ of bulk PE (~0.5 W/mK). Aligned polymer chains also yielded a high thermal conductivity of ~16 W/mK in polyethylene films drawn to large ratios approaching ~100.[8] More recently, the $k$ of aligned polyethylene-graphene nanocomposites was measured to be significantly higher compared to the $k$ of aligned PE.[7] While these studies shed light on $k$-enhancement through alignment effects at room temperature, experimental studies have also been performed to understand the temperature dependence of $k$ in aligned polymers. Choy *et al.*[9] measured the $k$ of aligned semi-crystalline polymers such as polypropylene and low-density polyethylene. $k$ along the alignment direction was found to increase monotonically with temperature up to the highest studied temperature of 300 K. More recently, Singh *et al.*[10] measured the $k$ of chain-oriented amorphous polythiophene and found it again to be a weakly increasing function of $T$. In this work, we use non-equilibrium molecular dynamics (MD) simulations[11,12] to study the temperature dependence of $k$ of aligned amorphous PE. Alignment is achieved through application of strain.

Molecular dynamics (MD) simulations have been used in the past to study thermal transport in polymers. Zhang and Luo[13] used MD simulations to study the temperature dependence of thermal conductivity of un-oriented amorphous PE and found it to reach a maximum at a temperature of 350 K. This peak in the $k$ of un-oriented amorphous PE was explained through morphological considerations by describing the increase in $k$ below 350 K in terms of increase in the radius of gyration and decrease in $k$ above 350 K in terms of reduced inter-chain interaction. Liu and Yang[14] used MD to investigate the role of strain rate in the $k$ of aligned amorphous PE at room temperature. Higher $k$-values were achieved for lower strain rates. Algaer *et al.*[15] used MD simulations to study the temperature dependence of $k$ of strained amorphous polystyrene; however, only small applied strains of up to 21% were considered in this study.

Aligned amorphous PE, the focus of study in this work, differs in several aspects compared to polymer systems discussed above, providing novel avenues to understand thermal transport in chain-oriented polymers. First, compared to semi-crystalline polymers, where two different phases exist (crystalline and amorphous), no such phase separation occurs in amorphous PE. Uniform morphology causes strain to uniformly impact the entire polymer structure. The absence of different phases can also eliminate phonon scattering that exists at the interface between crystalline and amorphous regions in semi-crystalline polymers, leading to potentially new features in the temperature dependence of $k$. Furthermore, while previous MD simulations of $k$ of amorphous polystyrene[15] considered strains of only up to 21%, in this work, we investigate much larger strains of 400%. At the small applied strains considered earlier, the change in the polymer structure was small, leading to a small change in the temperature dependence of $k$ with respect to the unstrained case. Much larger strain used in this work will lead to significant structural changes, enabling associated larger changes in the temperature dependence of $k$. Understanding the temperature dependence is of interest, both for gaining fundamental insights into the underlying thermal transport processes in strained polymers and for addressing low temperature applications of polymer based heat exchangers.[16,17]

## II. MOLECULAR DYNAMICS SIMULATIONS

To perform MD simulations of thermal transport, we use the LAMMPS[18] simulation package. Interatomic force interactions needed to compute atomic motion are modeled using the COMPASS force field.[19] COMPASS is an *ab initio* force field and has been used widely to simulate polymer systems. To study thermal transport in amorphous PE, the structure of the amorphous polymer is carefully constructed[13] (Fig. 1) by first equilibrating a single extended polyethylene chain of 1000 C atom length at 300 K for 1 ns to form a compact relaxed chain. The choice of 1000 C atom chain length leads to a good representation of the bulk behavior. 40 of these relaxed chains are then randomly packed into a cell leading to a total system size of $1.2 \times 10^5$ atoms. To achieve the final amorphous structure, the energy of the entire system is minimized followed by an increase in the system temperature to 600 K at a rate of 50 K/ns using an NPT (constant number of particles, pressure and temperature) ensemble, a further NPT run at 600 K for 4 ns to generate a polyethylene system with a relaxed and amorphous structure, then quenching the system to 300 K and finally an NPT run for 4 ns to equilibrate the structure at 300 K (Fig. 1).

After achieving the amorphous PE structure, deformation simulations were performed at 300 K to stretch the polymer. Strain was applied uniaxially along the x-axis of the periodic simulation cell to align the polymer chains along this direction (Fig. 1). Later, thermal conductivity is reported along the same direction in this work. Pressure was kept constant at 1 atm for all other boundaries during deformation using the NPT ensemble (Fig. 1). Strained polymer samples were further relaxed using NPT to obtain stable structures. The outlined procedure enabled polymer structures with strains of up to 400% to be achieved. The drawing process was performed at 300 K. To study the temperature dependence of $k$, the temperature of the strained polymer was varied between 50 and 400 K again using the NPT ensemble.

Thermal conductivity at different temperatures was computed using reverse non-equilibrium molecular dynamics (RNEMD) simulations based on the Muller-Plathe scheme[20] by imposing a heat flux $j_x$ across a simulation cell and estimating the resulting average temperature gradient, $\langle \partial T/\partial x \rangle$. Fourier's law of heat conduction, $k_x = -j_x/\langle \partial T/\partial x \rangle$, was then used to compute the thermal conductivity $k_x$ along the stretch direction. Through the Muller Plathe scheme, the system is divided into bins and heat flux is imposed by exchanging energies between the hot and cold bins located at the center and the edge of the system, respectively. To achieve a reliable estimate of the ensuing temperature gradient, RNEMD simulations were run for a time period of 1 ns, long enough to yield the steady state. Simulation was then performed for another 0.2 ns, during which the time-averaged temperature profile was estimated (typical profile for the unstretched polymer at 300 K is shown in Fig. 2). The linear section of this profile was used to compute the temperature gradient.

## III. RESULTS

The results of thermal conductivity simulation at different strains and temperatures are provided in Fig. 3 (error bars are estimated by repeating simulations with different starting configurations). First, we notice that increasing the strain from 0% (unstrained) to 400% causes a monotonic increase in thermal conductivity at each temperature. This enhancement in $k$ with increasing strain is well understood in terms of increasing alignment of the dominant heat conducting C-C covalent bonds in each polymer chain with the direction of heat transfer. Such polymer chain alignment, typically characterized by calculation of an average orientational order parameter[14] ($P_2$), is shown in Fig. 4 for different temperatures. $P_2$ is estimated by describing the local chain direction at each atom by a unit vector, $\mathbf{e}_i$, computed from the chord vectors connecting the atom to its nearest neighbors: $\mathbf{e}_i = (\mathbf{r}_{i+1} - \mathbf{r}_{i-1})/|\mathbf{r}_{i+1} - \mathbf{r}_{i-1}|$ and taking the projection of $\mathbf{e}_i$ along the alignment direction (x-direction) through the equation, $P_2 = 1.5 \langle (\mathbf{e}_i \cdot \mathbf{e}_x)^2 \rangle - 0.5$, where $\mathbf{e}_x$ is the unit vector in the direction of applied strain. The values of $P_2$ of the polymer samples after they are strained by different strains varying from 0% (unstretched) to 400% are shown in Fig. 4. For the unstretched case, the value of $P_2$ is close to 0 indicating randomly oriented chains. As the strain is increased, the value of $P_2$ increases, suggesting increasingly aligned chains. $P_2$ is also only weakly dependent on temperature, indicating that alignment is mostly a function of strain. Increase in $P_2$ leads to the observed increase in $k$ at each temperature.

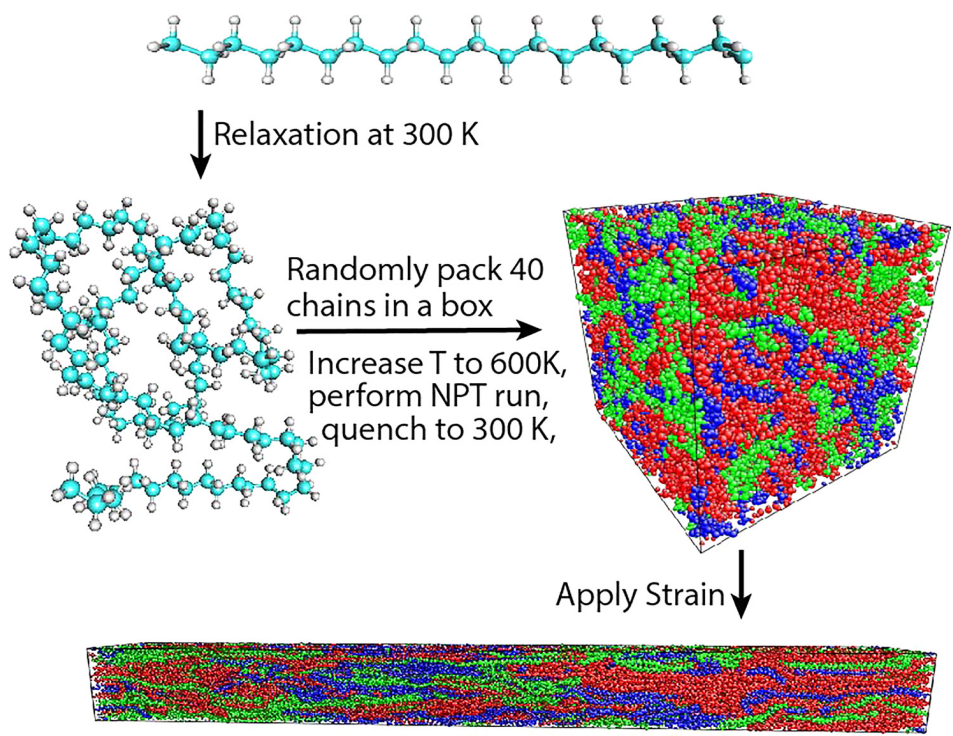

FIG. 1. Structure preparation of amorphous polyethylene and its alignment through application of strain.

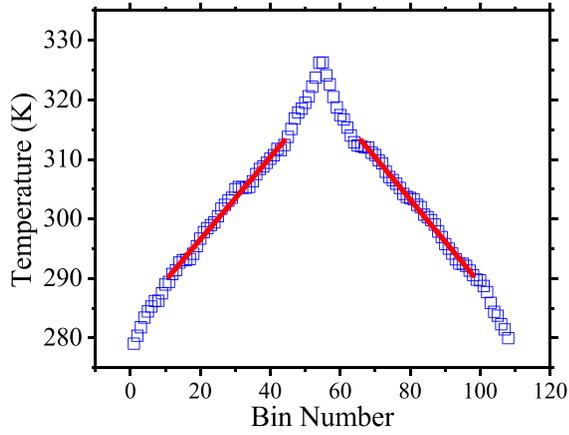

FIG. 2. Typical temperature profile obtained from NEMD simulations.

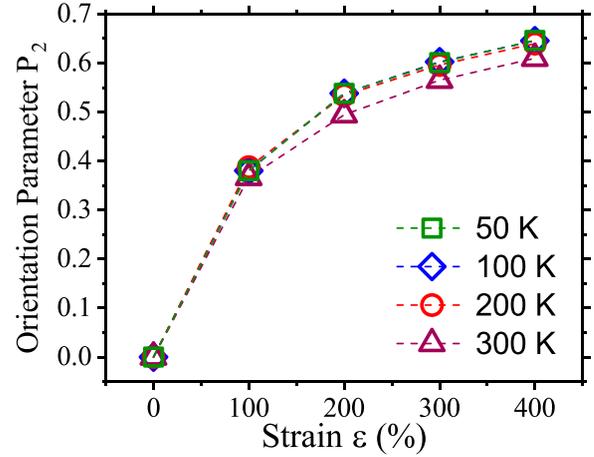

FIG. 4. Increase in the orientation parameter $P_2$ with strain at different temperatures.

We next investigate the effect of temperature on thermal conductivity. Figure 3 shows that the thermal conductivity for each strain reaches a maximum with temperature. For the unstretched polymer, thermal conductivity reaches a maximum at 350 K; this has been explained previously through morphological considerations.[13] The focus of this work is on studying the change in temperature dependence as the polymer is strained. Figure 3 shows that as the strain is increased, the temperature corresponding to maximum thermal conductivity ($T_{peak}$) shifts to lower values. For a strain of 100%, maximum thermal conductivity is reached at a temperature of 200 K. As the strain is further increased to 400%, the temperature corresponding to peak thermal conductivity decreases to 100 K. This temperature corresponding to peak $k$ for a strain of 400% is lower compared to that for the unstretched polymer by almost 250 K. The above results suggest that strained amorphous PE can be even more effective for thermal management at lower temperatures. This is seen by noticing that while at a temperature of 350 K, drawing the polymer to a strain of 400% leads to an increase in thermal conductivity from 0.30 W/mK (unstrained case) to 2.0 W/mK representing an enhancement of ~6.5-fold in $k$, at the lower temperature of 200 K, a much larger enhancement in $k$ of 18-fold is achieved from 0.24 W/mK to 4.43 W/mK, as the polymer is strained from 0% to 400%. We also studied the size effects by using a smaller system size of $0.6 \times 10^5$ atoms. $k$ was found to be higher for the larger system size of $1.2 \times 10^5$ atoms by 10%–15%; however, the peak $k$ (focus of this study) occurred at the same temperature for both these systems.

## IV. ROLE OF DISORDER AND ANHARMONICITY

The large shift in $T_{peak}$ with increasing strain can be understood in terms of an interplay between disorder and anharmonic phonon scattering. Heat in polymer systems is mainly conducted by lattice vibrations (phonons). Contribution of a phonon mode $\lambda$ to the overall thermal conductivity is described by[21] $k_\lambda \propto C_\lambda v_\lambda^2 \tau_\lambda$, where $C_\lambda$, $v_\lambda$ and $\tau_\lambda$ are the specific heat, the group velocity and the lifetime, respectively, of the phonon mode $\lambda$. The lifetime of a phonon mode is determined by the total scattering rate ($1/\tau_\lambda$), equal to the sum of scattering due to both disorder $(1/\tau_\lambda)^{disorder}$ and anharmonicity $(1/\tau_\lambda)^{\text{anharmonicity}}$ of polymer chains,[21] $1/\tau_\lambda = (1/\tau_\lambda)^{\text{disorder}} + (1/\tau_\lambda)^{\text{anharmonicity}}$. Disorder scattering involves the scattering of phonons from abrupt changes in the chain orientation and across polymer chains.[22,23] Disorder scattering rates are mostly temperature ($T$) independent, while anharmonic scattering rates increase linearly with temperature,[21] being weak at low $T$ and increasing at higher $T$. At low $T$, weak anharmonic effects result in disorder being the dominant scattering mechanism, while at higher $T$, anharmonic scattering increases in magnitude and begins to determine phonon lifetimes. $T_{peak}$ corresponds to this transition from disorder to anharmonic phonon transport. This is seen by observing that at low $T$, the dominant disorder scattering is independent of temperature,[21] causing $\tau_\lambda$ to become constant. Specific heat ($C$), however, has been known to increase with $T$ even in classical MD simulations due to anharmonic effects,[24] causing $k_\lambda$ to also increase with $T$ in the disorder dominated regime. At high $T$ in the anharmonicity dominated regime, however, anharmonic scattering rates increase with increasing $T$ causing phonon lifetimes, $\tau_\lambda$, to decrease as $T$ is increased, now causing the thermal conductivity ($k$) to decrease with increasing $T$. Increasing $k$ with increasing $T$ in the disorder dominated regime, followed by the opposite trend in the anharmonic regime, causes $k$ to reach a peak at the temperature corresponding to transition from disorder to anharmonicity dominated phonon transport.

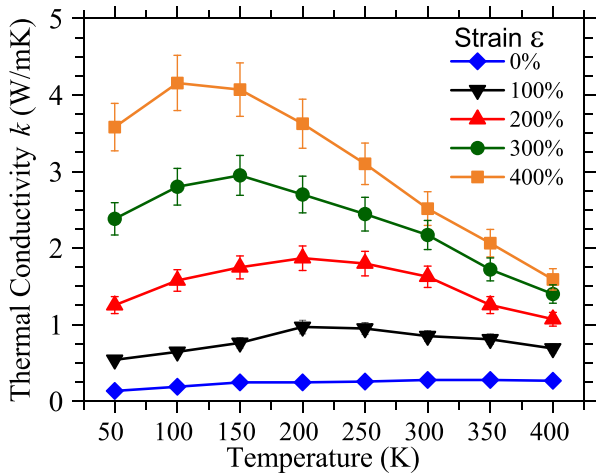

FIG. 3. Variation of thermal conductivity of amorphous polyethylene with temperature at different strains.

As the strain increases, transition from disorder to anharmonic regime shifts to lower $T$ causing $T_{peak}$ to also shift to lower values. This can be understood by noticing that at the low alignment levels achieved at low strains, polymer chains still have a large number of abrupt turns and bends, leading to a large disorder which causes disorder scattering rates to be large.[25] It requires a large increase in anharmonic scattering to exceed these large disorder scattering rates, which in turn requires high temperatures, thus causing the peak $k$ to occur at high $T$. As the applied strain is increased, the polymer structure becomes more aligned, causing the level of disorder to decrease. Relatively smaller increases in anharmonic scattering (requiring low $T$) can now cause it to overcome disorder scattering. Transition from disorder to the anharmonic regime thus occurs at progressively lower $T$ at higher strains, explaining the shift in peak $k$ to lower $T$ with increasing strain observed in Fig. 3.

We provide evidence supporting the above-presented mechanism for a shift in $T_{peak}$ with increasing strain, by modifying the system disorder and observing the resulting changes in $T_{peak}$. According to the presented mechanism, increasing the system disorder should shift $T_{peak}$ to higher values. At higher disorder, transition from disorder to anharmonicity dominated phonon transport would require higher anharmonic scattering rates, which in turn requires higher temperatures, causing $T_{peak}$ to shift to higher values. Vice versa, a decrease in disorder should shift the $T_{peak}$ to lower values. Both anharmonicity and disorder can be changed by modifying the potential parameters. The effect of change in disorder can, however, be studied in isolation without impacting other parameters such as phonon frequencies. This can be achieved by changing the dihedral energy parameters. Dihedral energy parameters control torsion angles, thus enabling control of the structure of the polymer chain, and therefore disorder. Since dihedral energies are much smaller than bond energies (as seen by noticing that the largest coefficients in energy terms for bond energy are of the order of 345 kcal/mol, while those for dihedral energies are ∼0.1 kcal/mol), changing dihedral terms does not significantly impact vibrational frequencies (which are instead determined by bond energies), enabling understanding the effect of change in disorder alone. Changing anharmonicity, however, requires a change in bond energy parameters (since bond energy terms are the largest contributors to anharmonicity), which also leads to a change in vibrational frequencies, precluding a study of the effect of change in anharmonicity alone. We therefore choose to study the effect of change in disorder through a change in dihedral energy parameters of the C-C-C-C dihedral, as these are most relevant for controlling the polymer backbone chain structure.

Through COMPASS[19] potential, the dihedral energy is computed using the expression $E_{dihedral} = K1(1 - \cos\phi) + K2(1 - \cos(2\phi)) + K3(1 - \cos(3\phi))$. For the C-C-C-C dihedral, the parameters $K1$, $K2$, and $K3$ are given by $K1 = 0$, $K2 = 0.054$ and $K3 = -0.143$ kcal/mol. The energy for these parameters is shown in Fig. 5 (solid red line) and the distribution of the dihedral angles for this case is shown in Fig. 6 (labeled "Original" in the figure). The dihedral angle of 180° corresponds to *trans* conformation, while dihedral angles of 60° and 300° correspond to *gauche* transformation. *Trans* conformations lead to stiff straight chains, while *gauche* transformations allow a change in the orientation, introducing disorder. By manipulating the parameters, $K1$, $K2$, and $K3$, the relative distribution of *trans* and *gauche* conformations can be modified, changing disorder.

As a first effect, we increase disorder by changing the parameters to $K1 = 0$, $K2 = -0.1$, and $K3 = -0.43$ kcal/mol for the polymer with 400% strain. This corresponds to the energy profile shown by the dashed blue line in Fig. 5. By lowering the energy corresponding to the *gauche state*, the number of gauche transformations is increased [Fig. 6(a)], thereby increasing disorder. To compute $k$ for the new set of parameters, a polymer matrix was prepared with these modified dihedral coefficients, strained by 400%, and finally relaxed to different temperatures to compute $k$ as a function of temperature. The results are shown in Fig. 7 (values for the original and modified dihedral parameters are shown by solid and open squares, respectively). First, it is noticed that compared to $k$ computed using original dihedral parameters (solid squares in Fig. 3), the $k$ computed using modified parameters is lower, as expected for a system with increased disorder. Second, while $k$ through the original parameters reached a peak at 100 K (Figs. 3 and 7), the $k$ using modified parameters (leading to higher disorder) reaches a maximum at a higher temperature of 150 K (Fig. 7). This is in agreement with the presented mechanism.

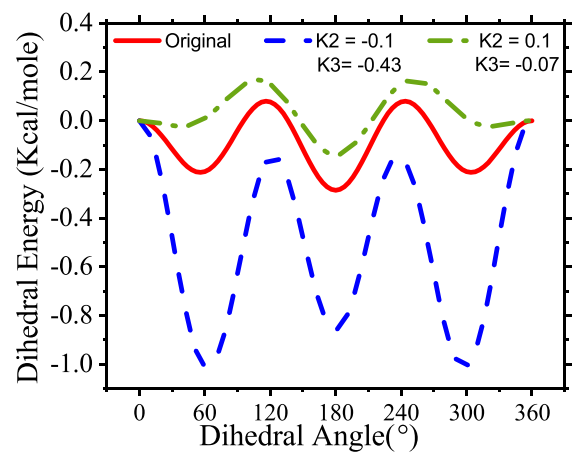

FIG. 5. Dihedral energy for the original COMPASS potential and for modified dihedral parameters.

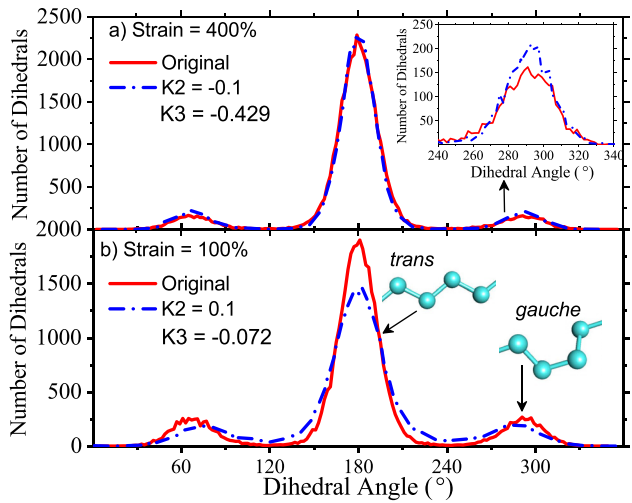

FIG. 6. Distribution of dihedral angles for modified and original dihedral parameters for strains of (a) 400% and (b) 100%.

Next, we also decrease disorder by manipulating the dihedral energy parameters and study its impact on $k$ of the polymer with a strain of 100%. A decrease in disorder should cause anharmonicity to become dominant at lower temperatures, shifting the peak $k$ to lower temperatures. To achieve lower disorder, we decrease the number of *gauche* transformations by using the parameters of $K1 = 0$, $K2 = 0.1$, and $K3 = -0.072$ kcal/mol. These parameters cause the energy corresponding to the *gauche* state to increase (dashed-dotted line in Fig. 5), causing the number of *gauche* transformations to decrease [Fig. 6(b)], thus lowering disorder. The thermal conductivity of the strained polymer with a strain of 100% and prepared with these new dihedral parameters is shown in Fig. 7 (as open triangles). The $k$ value for the new set of parameters is found to be higher compared to the original COMPASS potential, as expected for a system with lower disorder. Again, while $k$ through the original parameters reached the peak at 200 K, the $k$ using modified parameters (leading to lower disorder) reaches the maximum at a lower temperature of 150 K, again in agreement with the presented mechanism.

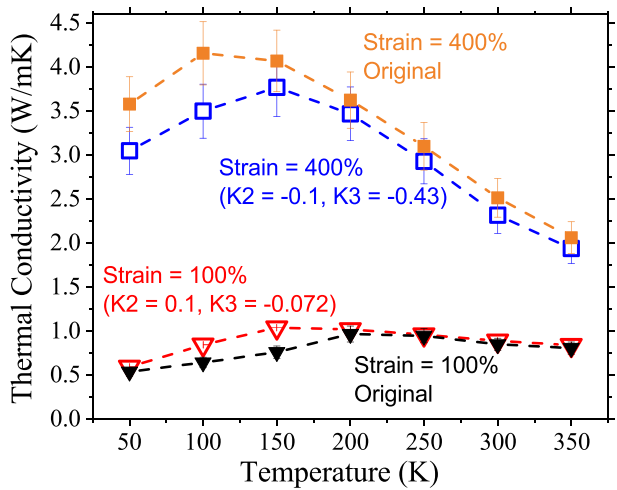

FIG. 7. Temperature dependence of $k$ of amorphous PE with modified and original dihedral parameters for strains of 100% and 400%.

The above results validate the presented mechanism for a shift in the $T_{peak}$ to lower values with increasing strain. The presented results for strained amorphous PE differ from measurements on strained amorphous polythiophene, where for the strained case, $k$ was found to be a weakly increasing function of $T$ even at higher temperatures.[10] This was explained in terms of presence of only short-range order in strained polythiophene, leading to phonon transport being dominated by disorder even in a strained system. The opposite trend of decreasing $k$ with increasing $T$ in strained amorphous PE at higher $T$ (∼200–300 K) suggests longer range order which causes phonon transport to be anharmonicity dominated at higher $T$. The above results, showing a transition of peak $k$ to lower temperatures, suggest that strained polymers can be even more effective for thermal management at lower temperatures.

## V. CONCLUSION

In summary, we have studied the temperature dependence (in the range of 50–400 K) of thermal conductivity of amorphous polyethylene drawn to strains of up to 400% using molecular dynamics simulations. The results demonstrate that the thermal conductivity ($k$) peaks with respect to temperature for all strains; this temperature corresponding to peak thermal conductivity shifts to lower values as the applied strain increases. While the $k$ of the unstretched polymer is maximum at 350 K, the $k$ of polymer with a strain of 100% peaks at 200 K. This values further decreases to 100 K for a strain of 400%. The effect is explained in terms of a cross-over from disorder to anharmonicity dominated phonon scattering regime. Increasing strain decreases disorder, allowing anharmonic scattering to become dominant at progressively lower temperatures, causing the peak $k$ temperature to shift to lower values. The effect is validated by modifying disorder through a change in dihedral energy parameters. Increasing the disorder is found to shift the temperature related to peak $k$ to higher values; vice versa, decreasing the disorder lowers that temperature. Both these results agree with the presented mechanism. A shift in peak $k$ to lower temperatures at higher strains leads to significant enhancement in the $k$ of aligned amorphous PE at lower temperatures. These results can lead to new avenues for the use of aligned polymers for thermal management at sub-ambient temperatures.